\documentclass[%
 reprint,
superscriptaddress,
showpacs,preprintnumbers,
 amsmath,amssymb,
 aps,
]{revtex4-1}

\usepackage{graphicx}
\usepackage{dcolumn}
\usepackage{bm}
\usepackage{color}

\begin{document}


\title{Specular Andreev reflection in thin films of topological insulators}
\author{Leyla Majidi}
\affiliation{School of Physics, Institute for Research in Fundamental Sciences (IPM), Tehran 19395-5531, Iran}
\author{Reza Asgari}
\affiliation{School of Physics, Institute for Research in Fundamental Sciences (IPM), Tehran 19395-5531, Iran}
\affiliation{Condensed Matter National Laboratory, Institute for Research in Fundamental Sciences (IPM), Tehran 19395-5531, Iran}

\date{\today}

\begin{abstract}
We theoretically reveal the possibility of specular Andreev reflection in a thin film topological insulator normal-superconductor (N/S) junction in the presence of a gate electric field. The probability of specular Andreev reflection increases with the electric field, and electron-hole conversion with unit efficiency happens in a wide experimentally accessible range of the electric field. We show that perfect specular Andreev reflection can occur for all angles of incidence with a particular excitation energy value. In addition, we find that the thermal conductance of the structure displays exponential dependence on the temperature. Our results reveal the potential of the proposed topological insulator thin-film-based N/S structure for the realization of intraband specular Andreev reflection.
\end{abstract}
\maketitle
\section{\label{sec:intro}Introduction}
Investigation of electron-transport properties of normal-superconductor
(N/S) hybrid nanostructures has attracted considerable attentions~\cite{exp}. Among the interesting effects is Andreev reflection (AR), through which an incident electron with excitation energy $\varepsilon$, upon hitting the N/S interface, is retro-reflected as a hole (with the same energy) retracing the same trajectory~\cite{Andreev64, linder15}. This peculiar scattering process provides a conversion of the dissipative electrical current in the N region into a dissipationless supercurrent and results in a finite conductance of a N/S junction at bias voltages below the superconducting gap~\cite{Blonder82}.

Recent studies have pointed out that novel interesting phenomena arise when N/S proximity structures are realized in atomically thin two-dimensional (2D) crystals. Beenakker showed that the peculiar band structure of graphene gives rise to the appearance of specular AR, which is absent in ordinary metal-superconductor interfaces~\cite{beenakker06,beenakker08}. In undoped graphene systems, from the band structure point of view, specular AR occurs where the electron-hole conversion is interband and the incident electron and the reflected hole are respectively from the conduction and valence bands~\cite{graphene-ex,graphene}. Recently, Lv and co-workers found the possibility of intraband specular AR in a corresponding structure with a 2D semiconductor in the presence of a strong Rashba spin-orbit coupling~\cite{Lv12}. They showed that, in the limit of low density or a strong spin-orbit coupling, specular AR is finite. Notice that strong Rashba spin-orbit interaction is rarely found in standard 2D crystal systems. Moreover, there are two prerequisites that must be satisfied in order to observe specular AR experimentally: the N/S interface should be transparent and well defined, and the N sample must be of high electronic quality~\cite{Calado}.

Thanks to the state-of-art semiconductor technologies, low-dimensional structures of three-dimensional (3D) topological insulators (TIs) can be routinely fabricated into ultrathin films~\cite{Qin09,zhang10} with the advantage that they have minimum bulk contribution. An ultrathin TI film is interesting when its thickness becomes comparable to the penetration depth of the helical surface states into the bulk, and top and bottom surfaces thus start to hybridize~\cite{zhang10}. The hybridization via quantum tunneling opens a gap in the band structure, which displays an oscillatory dependence on the thin film thickness~\cite{Linder09,Liu10,Lu10}. The variation of the thin film thickness can even lead to the change of the surface band Chern numbers and their topological properties~\cite{Lu10}.

In the present work, we investigate precisely the signature of the specular AR process by analyzing the electronic transport through a TI thin film N/S junction, as sketched in Fig. \ref{Fig:1}(a). We realize the possibility of generating intraband specular AR in the proposed N/S structure in the presence of a gate electric field. Within the scattering formalism, we find that increasing the magnitude of the gate-induced potential difference between the top and bottom surfaces, $2U$, of the TI leads to the enhancement of the probability of specular AR, and perfect specular electron-hole conversion occurs at near normal incidence to the N/S structure with different excitation energies, for a wide accessible experimental range of $U$. This is in contrast to the corresponding 2D electron gas (2DEG) structure~\cite{Lv12}, where perfect specular electron-hole conversion occurs in the presence of a strong Rashba spin-orbit interaction ($\lambda=0.4$ eV) only in the case of $\varepsilon=\Delta_S$ ($\Delta_S$ is the superconducting gap). Moreover, we show that this perfect specular AR happens for all angles of incidence to the proposed structure with $\varepsilon=\Delta_S$, when $U$ is large. This is another advantage of our proposed structure over the graphene- and 2DEG-based structures~\cite{beenakker06,beenakker08,Lv12}, where the electron-hole conversion with the unit efficiency occurs for normal incidence to the N/S interface. Then, we evaluate the Andreev differential conductance of the N/S structure and demonstrate that it increases with the subgap bias voltage $eV/\Delta_S$ for small values of $U$ and $\omega$, when $U=\omega$.
Also, we investigate the thermal transport characteristics of the proposed structure by evaluating the thermal conductance. We show that the thermal conductance increases exponentially with the temperature, which reflects the s-wave symmetry of the superconducting TI thin film.
\par
This paper is organized as follows. Section \ref{sec:level1} is devoted to the theoretical model and basic formalisms which will be used to investigate AR in a TI thin film N/S junction. In Sec. \ref{sec:level2}, we present our numerical results for the probabilities of the normal and Andreev reflection processes, the Andreev differential conductance, and the thermal conductance of the proposed structure. Finally, a brief summery of results is given in Sec. \ref{sec:level3}.
\section{\label{sec:level1}Model and Theory}
\begin{figure}[t]
\begin{center}
\includegraphics[width=3.4in]{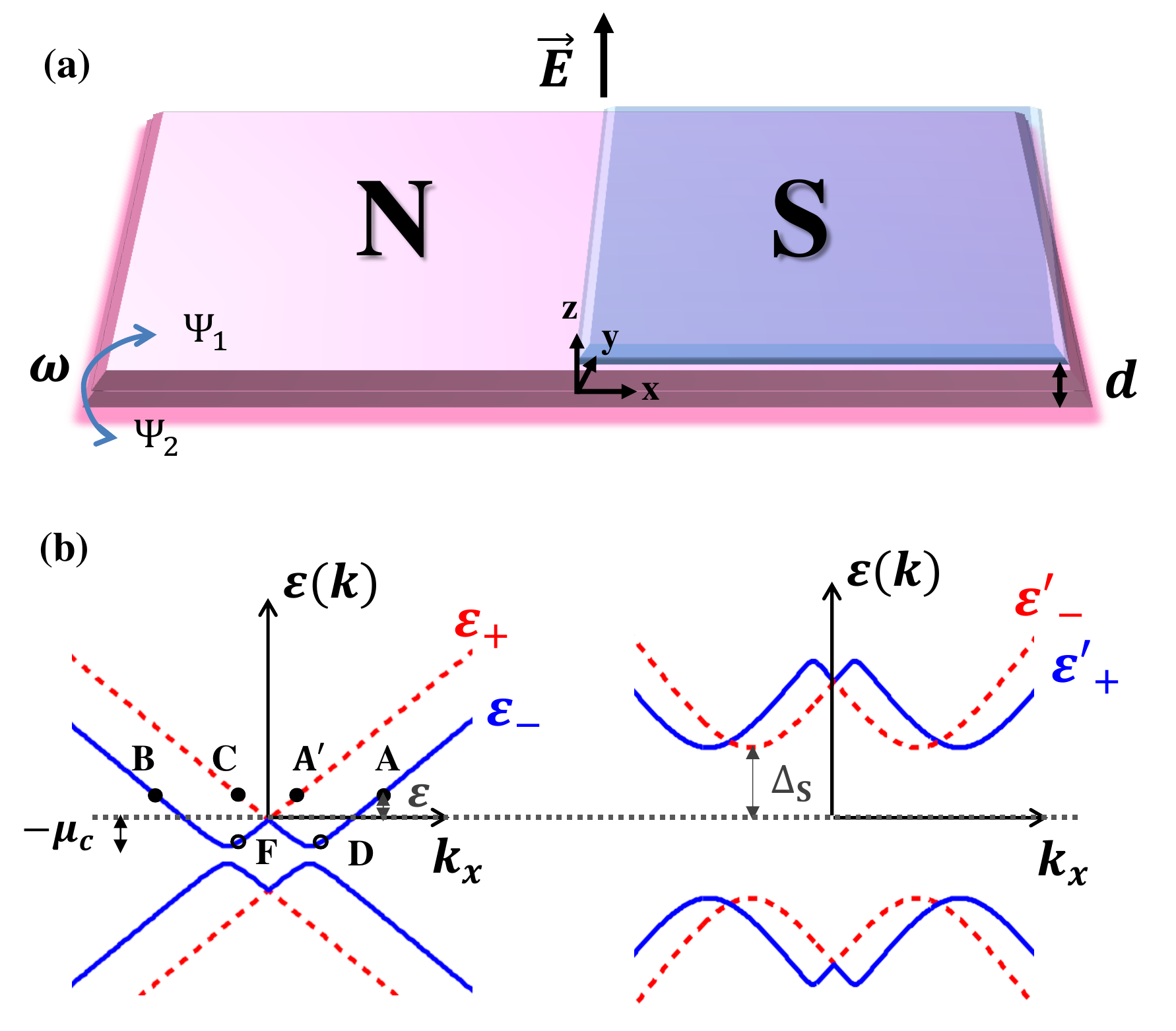}
\end{center}
\caption{\label{Fig:1}(a) Schematic illustration of the TI thin film N/S junction in presence of a perpendicular electric field $\bm{E}$, which can be realized by pairs of gate electrodes ($U_{top}=U$, $U_{bottom}=-U$). (b) The dispersion relation along the $x$ axis in momentum space [$\varepsilon=\varepsilon(k_x)$] of N (left panel) and S (right panel) regions, when $\omega=0.05$ eV, $U=0.2$ eV, $\mu_N=\mu_c$ ($\mu_c=\sqrt{U^2+\omega^2}$), $\mu_S=1$ eV, and $\Delta_S=0.5$ eV. An incident electron from the A (A') point ["-" ("+") branch of the spectrum] with excitation energy $\varepsilon>0$ can be normally reflected as an electron from B ("-" branch) and C ("+" branch) points and Andreev reflected as a hole from D ("-" branch) and F ("-" branch) points, with excitation energy $-\varepsilon$.}
\end{figure}
In order to study specular AR in a TI thin film in the presence of a perpendicular electric field, we consider the simplest experimental hybrid structure that can probe this phenomenon: a N/S junction as sketched in Fig. \ref{Fig:1}(a). The perpendicular electric field can be realized by pairs of gate electrodes ($U_{top}=U$, $U_{bottom}=-U$). The superconducting correlations can be induced in the TI thin film via the proximity effect to the S electrode with desired properties. Recently, proximity-induced superconductivity was experimentally demonstrated in a Bi$_2$Se$_3$ thin film by growing the TI on a conventional NbSe$_2$ superconductor~\cite{wang12, Finck}, and it was also addressed theoretically~\cite{fariborz}. To describe the superconducting correlations between electrons and holes with the wave functions $u$ and $v$, we use the following Dirac-Bogoliubov-de Gennes (DBdG) equation~\cite{linder10}
\par
\begin{equation}
\label{DBdG}
\hspace{-0.5cm}\left(
\begin{array}{cc}
\hat{H}(\bm{p})-\mu & \hat{\Delta}_S \\
-\hat{\Delta}^{\ast}_{S}& \mu-\hat{H}^{\ast}(\bm{-p})
\end{array}
\right)
\left(
\begin{array}{c}
u\\
v
\end{array}
\right)
=\varepsilon\left(
\begin{array}{c}
u\\
v
\end{array}
\right),
\end{equation}
with $\mu$ the chemical potential, $\hat{H}(\bm{p})$ the single-particle Hamiltonian, and $\hat{\Delta}_S$ the superconducting pair potential which couples the time-reversed electron and hole states.
\par
The effective single-particle Hamiltonian of a TI thin film~\cite{Zyuzin11,Pershoguba12} in the presence of a perpendicular gate electric field has the form
\begin{equation}
\label{H}
\hat{H}(\bm{p})=\hat{{\tau}}_z\otimes \hat{h}(\bm{p})+\hat{\tau}_x\otimes\ \omega\ \hat{\sigma}_0+U\ \hat{\tau}_z\otimes\hat{\sigma}_0,\\
\end{equation}
which acts on a four-dimensional spinor $(\Psi_{1\uparrow},\Psi_{1\downarrow},\Psi_{2\uparrow},\Psi_{2\downarrow})$ . The indices 1,2 label the two surface states localized at opposite surfaces of the film with width $d$, as shown in Fig. \ref{Fig:1}(a). The 2D Hamiltonian $\hat{h}(\bm{p})$ in the subspace of real spin can be described by
\begin{eqnarray}
\hat{h}(\bm{p})=v_{\rm F} \hat{z}(\hat{\bm{\sigma}}\times\bm{p}),
\end{eqnarray}
where the terms up to linear in $\bm{p}$ are kept in the low-energy regime. It has been demonstrated that the presence of the quadratic term of $\bm{p}$ in the Hamiltonian $\hat{h}(\bm{p})$ together with the momentum dependence of the gap~\cite{Lu10,Linder09,Liu10} does not change results quantitatively~\cite{Parhizgar}. Therefore, we just consider the leading term in momentum of the low-energy Hamiltonian in this article since we consider the low-energy excitation energy.

While the wave functions of Eq. (\ref{H}) are localized in the $z$ direction, electrons are free to move parallel to the surface (in the $x$-$y$ plane). The wave functions of the two surface states $\Psi_1$ and $\Psi_2$ decay into the bulk and have a finite decay length $\xi$. So, when the thickness of the film becomes comparable with the decay length, $d\sim\xi$, there is a finite coupling between the surface states, $\omega$, which is assumed to be proportional to the unit matrix in the real spin. The two sets of Pauli matrices, $\hat{\sigma}$ and $\hat{\tau}$, act on the real spin and the top and bottom surface pseudospin degrees of freedom, correspondingly, $\hat{z}$ shows the direction perpendicular to the upper surface, $v_{\rm{F}}$ is the Fermi velocity of the helical states in the TI, and $2U$ is the potential difference between the top and bottom surfaces induced by the gate electric field. The pairing symmetry of the superconductor is assumed to be an $s$ wave, and the superconducting gap matrix $\hat{\Delta}_S$ is given as $\hat{\Delta}_S=i{\Delta}_S\hat{\sigma}_y\Theta(x)$, where $\Theta(x)$ is the Heaviside step function.
\par
 The uncoupled top and bottom surfaces of the TI thin film ($\omega=0$) exhibit linear dispersion relations with opposite momentum shifts ($\pm U$) in the presence of the gate electric field. On the other hand, by solving the DBdG equation for the TI thin film with coupled surfaces ($\omega\neq 0$), we directly find the energy dispersion for the N region with the chemical potential $\mu_N$ as
\begin{equation}
\label{E_N}
\varepsilon_{\pm}(\bm{k})=\pm\sqrt{(\hbar v_{\rm F}|\bm{k}|\pm U)^2+\omega^2}-\mu_N,
\end{equation}
where $\pm$ in front of the square root denotes the energy dispersion of the conduction/valence band and the energy gap is determined by the tunneling element $\omega$, and for the S region with the chemical potential $\mu_S$ as
\begin{equation}
\label{E_S}
\varepsilon'_{\pm}(\bm{k}_S)=\sqrt{(\mu_S-\sqrt{(\hbar v_{\rm F}|\bm{k}_S|\mp U)^2+\omega^2})^2+\Delta_S^2},
\end{equation}
 such that they are not the eigenenergies of the two surfaces, and are shown in Fig. \ref{Fig:1}(b) for the N region with $\mu_N=\mu_c$ (left panel) and for the highly doped S region with $\mu_S=1$ eV and $\Delta_S=0.5$ eV (right panel), when $\omega=0.05$ eV and $U=0.2$ eV. Note that we have used the large value of the superconducting gap $\Delta_S$ in order to clarify the plot of the energy dispersion in the S region.

It is known that a TI thin film  in the absence of the gate electric field can be described using two degenerate massive Dirac hyperbolas (with an energy gap of $2\omega$ between the conduction and valence bands) in the momentum space, which are each other's time-reversal counterpart. The gate-induced potential difference between the top and bottom surfaces leads to opposite momentum shifts ($\pm U$) on the two degenerate hyperbolas of the opposite surfaces, which are located at $z=\pm d/2$. Inducing the superconducting pair potential $\Delta_S$ to the highly doped TI thin film via the proximity to the S electrode leads to the coupling of the time-reversed electron and hole states and opening of a gap $2\Delta_S$ at the crossing point of the electron-like and hole-like spectra (zero energy). This can be seen from the right panel of Fig. \ref{Fig:1}(b).
\par
To investigate specular AR at the N/S interface, we consider the case that the chemical potential of the N region, $\mu_N$, lies at the crossing point of two branches of the electron spectrum with the chemical potential $\mu_c=\sqrt{U^2+\omega^2}$. At a certain energy $\mu_N+\varepsilon$, which lies slightly above the band crossing, there are two incident electron states with the wave vectors of A and A' points. An incident electron of the conduction-band from the left to the N/S interface, with a subgap energy $\varepsilon\leq\Delta_S$, can be either normally reflected as an electron or Andreev reflected as a hole in the same band.

As illustrated in the left panel of Fig. \ref{Fig:1}(b), there are four possible reflection modes for the incident electron from the A (A') point. Two of them are reflected conduction-band electrons with wave vectors at B and C points (respectively from "-" and "+" branches of the spectrum) and the other two are the reflected conduction-band holes with wave vectors at D and F points ("-" branch of the spectrum). By simple inspection of the energy dispersion [Fig. \ref{Fig:1}(b)] and considering the conservation of the transverse wave vector $k_y$ upon reflection at the interface $x=0$, we find that the incident electron from the A (A') point with wave vector $\bm{k}_{A(A')}=|\bm{k}_{A(A')}|(\cos{\alpha_{A(A')}},\sin{\alpha_{A(A')}})$ and group velocity $\bm{v}_{g,A(A')}={v}_{g,A(A')}(\cos{\alpha_{A(A')}},\sin{\alpha_{A(A')}})$ [obtained through $\bm{v_g}=\hbar^{-1}\nabla_{\bm{k}}\varepsilon(\bm{k})$] is specularly reflected as an electron with wave vector $\bm{k}_{B(C)}=|\bm{k}_{B(C)}|(-\cos{\alpha_{B(C)}},\sin{\alpha_{B(C)}})$ and $\bm{v}_{g,B(C)}={v}_{g,B(C)}(-\cos{\alpha_{B(C)}},\sin{\alpha_{B(C)}})$, so that ${\bm{k}_{B(C)}}.{\bm{v}_{g,B(C)}}$ has the same sign as that of the incident electron (${\bm{k}_{A(A')}}.{\bm{v}_{g,A(A')}}>0$).

Since a conduction-band hole moves opposite to its wave vector, the reflected hole from the D point with $\bm{k}_{D}=|\bm{k}_{D}|(\cos{\alpha_{D}},\sin{\alpha_{D}})$ retraces the path of the incident electron with $\bm{v}_{g,D}=-{v}_{g,D}(\cos{\alpha_{D}},\sin{\alpha_{D}})$ and therefore a retro type AR occurs at the D point with the opposite sign of $\bm{k}_{D}.\bm{v}_{g,D}$ to that of the incident electron from the A or A' point ($\bm{k}_{D}.\bm{v}_{g,D}<0$). Interestingly, it is found that, for the reflected conduction-band hole from the F point with $\bm{k}_{F}=|\bm{k}_{F}|(-\cos{\alpha_{F}},\sin{\alpha_{F}})$, the AR process will be specular with $\bm{v}_{g,F}={v}_{g,F}(-\cos{\alpha_{F}},\sin{\alpha_{F}})$ and $\bm{k}_{F}.\bm{v}_{g,F}>0$. The magnitude of the group velocity for the hole at D or F point is $v_{g,i}=v_{\rm F} |\hbar v_{\rm F}|\bm{k}_i|-U|/\sqrt{(\hbar v_{\rm F}|\bm{k}_i|-U)^2+\omega^2}$ ($i=D, F$), for the electron from the "-" branch of the spectrum it is $v_{g,i}=v_{\rm F} (\hbar v_{\rm F}|\bm{k_i}|- U)/\sqrt{(\hbar v_{\rm F}|\bm{k}_i|- U)^2+\omega^2}$ ($i=A, B$), and for the electron from the "+" branch of the spectrum it is $v_{g,i}=v_{\rm F} (\hbar v_{\rm F}|\bm{k}_i|+ U)/\sqrt{(\hbar v_{\rm F}|\bm{k}_i|+ U)^2+\omega^2}$ ($i=A', C$).

Therefore, stemming from aforementioned discussions, by ray analysis~\cite{Landau}, the AR with the wave vector at the D point is a retro-reflection, and that with the wave vector at the F point is specular reflection. We emphasize that the specular AR in the proposed N/S structure is intraband and both the incident electron and the reflected hole are from the same band (conduction band), while in the case of undoped graphene the specular AR occurs when the incident electron and the reflected hole are respectively from the conduction and valence bands~\cite{beenakker06,beenakker08}.

We note that, when the chemical potential of the N region lies above the band crossing point ($\mu_N>\mu_c$), the specular AR occurs the excitation energies $\varepsilon>\mu_N-\mu_c$, while in the case of $\mu_N<\mu_c$ the specular electron-hole conversion is possible for incident electrons from two electron states (A and A' points), when $\varepsilon>\mu_c-\mu_N$ and for only one electron state (the A point), when $\varepsilon<\mu_c-\mu_N$.
\par
Denoting the amplitudes of the normal and Andreev reflection processes, $r_{e,B}^{A(A')}$, $r_{e,C}^{A(A')}$, $r_{h,D}^{A(A')}$, and $r_{h,F}^{A(A')}$, respectively, the total wave function inside the N region can be written as
\begin{widetext}
\begin{eqnarray}
\label{N}
&&\psi_{N}^{(')}=A_{A(A')}^e\ e^{ik_{A^{(')},x}x} e^{ik_yy}
\left(
\begin{array}{c}
1\\
-i\ a_{A(A')}^{e}\ e^{i\alpha_{A(A')}}\\
b_{A(A')}^{e}\\
-i\ c_{A(A')}^{e}\ e^{i\alpha_{A(A')}}\\
0\\
0\\
0\\
0
\end{array}\right)+r_{e,B}^{A(A')} A_{B}^e\ e^{-ik_{B,x}x} e^{ik_yy}
\left(
\begin{array}{c}
1\\
i\ a_{B}^{e}\ e^{-i\alpha_{B}}\\
b_{B}^{e}\\
i\ c_{B}^{e}\ e^{-i\alpha_{B}}\\
0\\
0\\
0\\
0
\end{array}
\right)\nonumber\\
\nonumber\\
&&+r_{e,C}^{A(A')} A_{C}^e\ e^{-ik_{C,x}x} e^{ik_yy}
\left(
\begin{array}{c}
1\\
i\ a_{C}^{e}\ e^{-i\alpha_{C}}\\
b_{C}^{e}\\
i\ c_{C}^{e}\ e^{-i\alpha_{C}}\\
0\\
0\\
0\\
0
\end{array}
\right)+r_{h,D}^{A(A')} A_{D}^h\ e^{ik_{D,x}x} e^{ik_yy}
\left(
\begin{array}{c}
0\\
0\\
0\\
0\\
1\\
-i\ a_{D}^{h}\ e^{i\alpha_{D}}\\
b_{D}^{h}\\
-i\ c_{D}^{h}\ e^{i\alpha_{D}}
\end{array}
\right)+r_{h,F}^{A(A')} A_{F}^h\ e^{-ik_{F,x}x} e^{ik_yy}
\left(
\begin{array}{c}
0\\
0\\
0\\
0\\
1\\
i\ a_{F}^{h}\ e^{-i\alpha_{F}}\\
b_{F}^{h}\\
i\ c_{F}^{h}\ e^{-i\alpha_{F}}
\end{array}
\right),\nonumber\\\nonumber\\
\end{eqnarray}
for incoming electron from the A(A') point with the wave vector $\bm{k}_{A(A')}$. Here, $\alpha_i=\arcsin({k_y/|\bm{k_i}|})$ ($i=A, A', B, C, D, F$) indicates the angle of the propagation of the electron or the hole at a transverse wave vector $k_y$ with longitudinal wave vector $k_{i,x}=\sqrt{|\bm{k}_i|^2-k_y^2}$, such that $|\bm{k}_{A(A')}|=(\sqrt{(\mu_N+\varepsilon)^2-\omega^2}\pm U)/\hbar v_{\rm{F}}$, $|\bm{k}_B|=|\bm{k}_A|$, $|\bm{k}_C|=|\bm{k}_{A'}|$, and $|\bm{k}_{D(F)}|=U\pm\sqrt{(\mu_N-\varepsilon)^2-\omega^2}$. Also, we have defined $b_{i}^{e(h)}=[(\hbar v_{\rm{F}} |\bm{k}_i|)^2+\omega^2-(U-\mu_N\mp\varepsilon)^2]/2\omega U$, $c_{i}^{e(h)}=[\omega-b_{i}^{e(h)} (U+\mu_N\pm\varepsilon)]/\hbar v_{\rm{F}} |\bm{k}_i|$, $a_{i}^{e(h)}=[\hbar v_{\rm{F}} |\bm{k}_i| b_{i}^{e(h)}+c_{i,}^{e(h)} (U+\mu_N\pm\varepsilon)]/\omega$, and $A_{i}^{e(h)}=1/[\sqrt{(a_{i}^{e(h)}-b_{i}^{e(h)}c_{i}^{e(h)}) \cos{\alpha_i}}]$.
\par
Inside the S region, the total wave function is a superposition of two electron-like and two hole-like quasiparticle states
\begin{equation}
\psi_{S}=\Sigma_{l=1}^2\ t_l\ e^{ik_l x} e^{ik_y y}
\left(
\begin{array}{c}
1\\
A_l\\
B_l\\
C_l\\
D_l\\
F_l\\
G_l\\
H_l
\end{array}
\right)+\Sigma_{j=3}^4\ t_j\ e^{-ik_j x} e^{ik_y y}
\left(
\begin{array}{c}
1\\
A_j\\
B_j\\
C_j\\
D_j\\
F_j\\
G_j\\
H_j
\end{array}
\right),\\
\\
\end{equation}
\end{widetext}
from $\varepsilon'_+$ and $\varepsilon'_-$ branches of the spectrum, respectively, with the longitudinal wave vectors $k_l=k_{0l}+i\kappa_l$ and $k_j=k_{0j}-i\kappa_j$, which can be obtained using Eq. (\ref{E_S}) (where $|\bm{k_S}|$ is replaced with $k_S$) and $k_{l(j)}=\sqrt{{k^2_{S,l(j)}}-k_y^2}$. The other parameters of the quasiparticle states are defined by
\begin{eqnarray}
A_{l(j)}&=&\frac{M_3^2 M_8 M_9-N_{l(j)} P_{l(j)}-M_{2{l(j)}} M_{4{l(j)}}M_{10}M_{11}}{M_{2{l(j)}}(P_{l(j)} M_{11}+N_{l(j)} M_{10})},\nonumber\\
B_{l(j)}&=&\frac{N_{l(j)}+M_{2{l(j)}}A_{l(j)}M_{11}}{M_3M_8},\nonumber\\
C_{l(j)}&=&\frac{M_{4{l(j)}}{M_{11}+N_{l(j)} A_{l(j)}}}{M_3M_8},\nonumber
\end{eqnarray}
$D_{l(j)}=M_{4{l(j)}}+M_1 A_{l(j)}+M_3 C_{l(j)}$, $F_{l(j)}=-M_1-M_{2{l(j)}} A_{l(j)}-M_3 B_{l(j)}$, $G_{l(j)}=M_3 A_{l(j)}-M_{4{l(j)}} B_{l(j)}-M_5 C_{l(j)}$, $H_{l(j)}=-M_3+M_5 B_{l(j)}+M_{2{l(j)}} C_{l(j)}$, where $N_{l(j)}=M_{1} M_{6} +M_{2{l(j)}} M_{4{l(j)}}+M_3^2+1$, $P_{l(j)}=M_{5} M_{7} +M_{2{l(j)}} M_{4{l(j)}}+M_3^2+1$, $M_{1(5)}=[U\mp(\mu_S+\varepsilon)]/\Delta_S$, $M_{2{l(j)}}=M_{4{l(j)}}^{\ast}=(k_y+ik_{l(j)})/\Delta_S$, $M_3=\omega/\Delta_S$, $M_{6(7)}=[U\mp(\mu_S-\varepsilon)]/\Delta_S$, $M_8=M_9=2\mu_S/\Delta_S$, and $M_{10(11)}=2(U\pm\mu_S)/\Delta_S$.
\par
Matching the wave functions of N and S regions at the interface $x=0$, the scattering coefficients for the normal and Andreev reflection processes can be obtained. Therefore, we can calculate the Andreev differential conductance of the TI thin-film-based N/S structure of width $W$ by using the Blonder-Tinkham-Klapwijk (BTK) formula~\cite{Blonder82} at zero temperature,
\begin{eqnarray}
\label{G_AR}
G(eV)&=&\frac{e^2 W}{2\pi^2 \hbar}\sum_{i=A,A'}\int_{0}^{|\bm{k}_i(eV)|}[1-(|r_{e,B}^i|^2+|r_{e,C}^i|^2)\nonumber\\
&+&(|r_{h,D}^i|^2+|r_{h,F}^i|^2)]\ dk_y,
\end{eqnarray}
where the summation is over two possible incident electron states and we put $\varepsilon=eV$ at zero temperature.

Having known the reflection coefficients, we can calculate the thermal transport properties of the N/S structure. Assuming a temperature gradient $\Delta T$ through the junction, we can further evaluate the thermal conductance $\kappa=\lim_{\Delta T\to 0}J_{th}/\Delta T$, with $J_{th}$ the heat current density, by incorporating the low-energy excitations as follows ~\cite{Bardas95,Yokoyama08}:
\begin{eqnarray}
\label{kappa}
\kappa&=&\frac{k_B W}{8\pi^2 \hbar}\sum_{i=A,A'}\int_0^{\infty} \int_{0}^{|\bm{k}_i(\varepsilon)|}d\varepsilon\ dk_y\  \frac{\varepsilon^2}{(k_B T)^2{\cosh^2(\frac{\varepsilon}{2k_B T})}}\nonumber\\
&\times&\{1-[|r_{e,B}^i|^2+|r_{e,C}^i|^2]-[Re(\frac{\cos{\alpha_D}}{\cos{\alpha_i}})\ |r_{h,D}^i|^2\nonumber\\
&+&Re(\frac{\cos{\alpha_F}}{\cos{\alpha_i}})\ |r_{h,F}^i|^2]\},
\end{eqnarray}
where we replace the zero-temperature superconducting order parameter $\Delta_S$ in Eq. (\ref{DBdG}) with the temperature-dependent one, $\Delta_S(T)=1.76 k_B T_C\tanh{(1.74\sqrt{{T_C}/{T}-1})}$.

\section{\label{sec:level2}NUMERICAL RESULTS and DISCUSSIONS}
\begin{figure}[t]
\begin{center}
\includegraphics[width=3.4in]{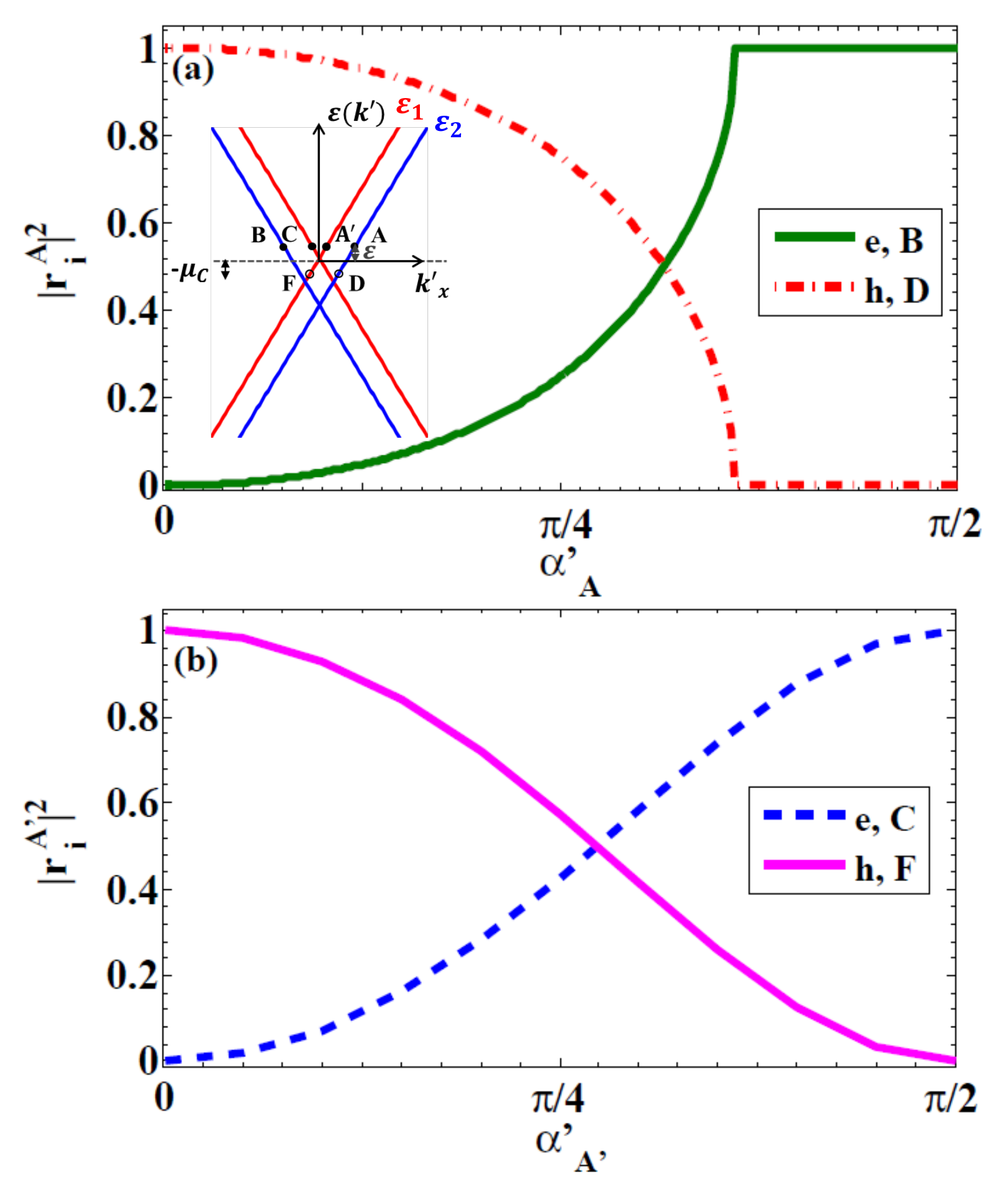}
\end{center}
\caption{\label{Fig:2} The normal and Andreev reflection probabilities for an incident electron from (a) A and (b) A' points versus the angle of incidence, when there is no coupling between the top and bottom surfaces of the TI thin film ($\omega=0$), $U=0.05$ eV, and $\varepsilon/ \Delta_S = 0.5$. The superscript $e, B(C)$ denotes the normal reflection of the electron from the B (C) point, and $h,D(F)$ denotes the retro (specular) AR from the D (F) point. The inset of (a) shows the dispersion relation (along the $x$ axis in momentum space) of the uncoupled top (red lines) and bottom (blue lines) surfaces of the N region. An incident electron from A (A') point [bottom (top) surface] can be normally reflected as an electron from the B (C) point and Andreev reflected as a hole from the D (F) point.}
\end{figure}
\begin{figure}[t]
\begin{center}
\includegraphics[width=3.4in]{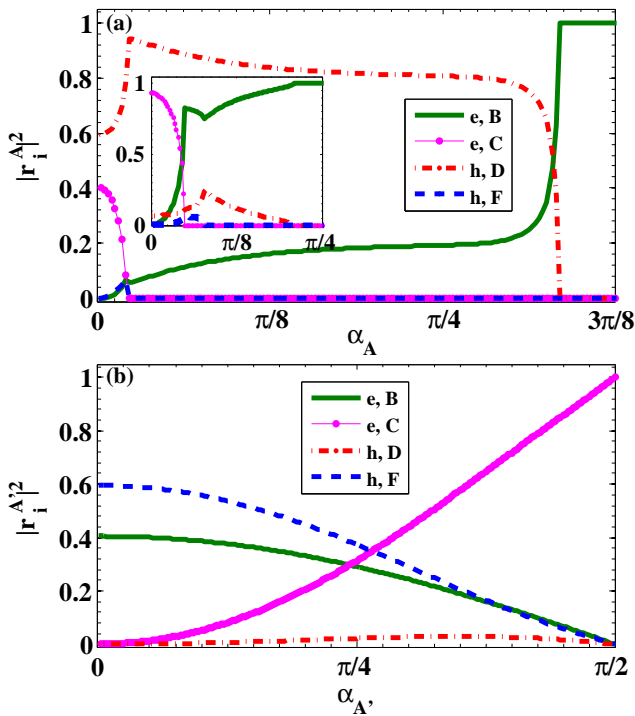}
\end{center}
\caption{\label{Fig:3} Probabilities of the normal and Andreev reflection processes for four possible scattering modes of an incident electron from (a) A and (b) A' points to the TI thin-film-based N/S interface versus the angle of incidence, when $\omega=0.05$ eV, $U=\omega$, and $\varepsilon/ \Delta_S = 0.5$. The inset of (a) shows the normal and Andreev reflection probabilities for the incident electron from the A point, when $\omega=0.2$ eV.}
\end{figure}
To evaluate the numerical results, using the numerical reflection amplitudes $r_{e,B}^{A(A')}$, $r_{e,C}^{A(A')}$, $r_{h,D}^{A(A')}$, $r_{h,F}^{A(A')}$, and Eqs. (\ref{G_AR}) and (\ref{kappa}), we set the Fermi velocity $v_{\rm{F}}=5\times 10^5$ m/s, the Boltzmann constant $k_B=1$, the zero-temperature superconducting energy gap $\Delta_S = 0.01$ eV, and the top gate potential in a range $U=0.05-0.4$ eV. The tunneling element $\omega$ between the top and bottom surfaces  depends experimentally on the width of the thin film and varies from $0.25$ eV for the ultrathin $2$ nm film to $0.05$ eV for the $5$ nm film of Bi$_2$Se$_3$~\cite{zhang10}. Also, we set the chemical potential of the N region to lie on the crossing point of the two branches of the spectrum $\mu_N=\mu_c$ ($\mu_c=\sqrt{U^2+\omega^2}$) and the chemical potential of the S region $\mu_S = 1$ eV, since the S region is often heavily doped in experimental situations. Furthermore, we scale the temperature, T, in units of the critical temperature of the superconducting order parameter, $T_C$, and the excitation energy, $\varepsilon$, in units of the superconducting energy gap, $\Delta_S$. To check out our numerical calculations, we examine the graphene set-up considering $U\rightarrow 0$ and $\omega \rightarrow 0$.
\par
First, we plot the probability of the normal and Andreev reflection processes for an incident electron from the A (A') point in terms of the angle of incidence $\alpha'_A$ ($\alpha'_{A'}$) in Fig. \ref{Fig:2}(a) [Fig. \ref{Fig:2}(b)], when there is no coupling between the top and bottom surfaces of the TI thin film ($\omega=0$), $U =0.05$ eV, and $\varepsilon/\Delta_S=0.5$. As shown in the inset of Fig. \ref{Fig:2}(a), the uncoupled top and bottom surfaces of the N region exhibit linear dispersion relations $\varepsilon_1(|\bm{k}'|)=\pm \hbar v_{\rm F}|\bm{k}'|$ and $\varepsilon_2(|\bm{k}'|)=\pm \hbar v_{\rm F}|\bm{k}'|-2U$, respectively for the top (red lines) and bottom (blue lines) surfaces in the presence of the applied gate electric field. We note that $\varepsilon_1$ and $\varepsilon_2$ are the eigenenergies of the effective single-particle Hamiltonian of the TI thin film [Eq. (\ref{H})] in the absence of the coupling parameter between the two surfaces, and they can-not be obtained from Eq. (\ref{E_N}) by setting $\omega=0$, since the $\varepsilon_+$ and $\varepsilon_-$ defined in Eq. (\ref{E_N}) are not the eigenenergies of the top and bottom surfaces.
By similar ray analysis, it is found that the incident electron from the top surface (red line) with wave vector $\bm{k}'_{A'}$ and group velocity $\bm{v}'_{g,A'}$ (with the $\bm{k}'_{A'}.\bm{v}'_{g,A'}>0$) is normally reflected as an electron at the C point and Andreev reflected as a hole at the F point via the interband specular type AR, since the conduction-band electron is reflected as a valence-band hole with $\bm{k}'_{F}.\bm{v}'_{g,F}>0$. In the case of an incident electron from the bottom surface (blue line) with wave vector $\bm{k}'_A$ and  $\bm{k}'_{A}.\bm{v}'_{g,A}>0$, normal reflection takes place at the B point and retro type AR occurs for the conduction-band hole at the D point with the $\bm{k}'_{D}.\bm{v}'_{g,D}<0$. The magnitude of the group velocity for the electron or the hole is $|\bm{v}'_{g,i}|=v_{\rm F }$ ($i=A,B,C,D,F$). It is shown in Fig. \ref{Fig:2}(a) that in the case of an incident electron with wave vector $\bm{k}'_A$ (from the bottom surface), the probability of retro type electron-hole conversion ,$|r_{h,D}^A|^2$, decreases by increasing $\alpha'_A$ from unit efficiency at normal incidence and goes to zero for $\alpha'_A>\arcsin(|\bm{k}'_D|/|\bm{k}'_A)$, while the normal reflection probability, $|r_{e,B}^A|^2$, increases with $\alpha'_A$ and reaches unity when the Andreev reflected hole becomes evanescent and does not contribute to any transport of charge.

In the case of an incident electron with wave vector $\bm{k}'_{A'}$ (from the top surface), it is demonstrated in Fig. \ref{Fig:2}(b) that the probability of normal reflection from the C point, $|r_{e,C}^{A'}|^2$, has an increasing behavior with $\alpha'_{A'}$, and perfect normal reflection occurs at $\alpha'_{A'}=\pi/2$, while the probability of specular AR from the F point, $|r_{h,F}^{A'}|^2$, decreases from unity at normal incidence and tends to zero for $\alpha'_{A'}=\pi/2$. We note that, from the band structure point of view, this specular type AR is interband, in which the incident electron from the conduction band (A' point) is Andreev reflected as a hole in the valence band (F point), while, in the case of intraband specular AR in the thin-film-based structure ($\omega\neq 0$), both the incident electron and the reflected hole are from the same band (conduction band).

Now, with this knowledge, we turn on the coupling parameter $\omega$. The hybridization of the top and bottom surfaces of the TI thin film leads to the contribution of two more scattering processes: a normal reflection from the C point and an Andreev reflection from the F point. Considering these two reflection processes, we present the behavior of the probability of normal ($|r_{e,B}^{A}|^2, |r_{e,C}^{A}|^2$) and Andreev ($|r_{h,D}^{A}|^2, |r_{h,F}^{A}|^2$) reflection processes as a function of $\alpha_A$ for an incident electron from the A point in Fig. \ref{Fig:3}(a), when the coupling parameter $\omega=0.05$ eV, $U=\omega$, and $\varepsilon/ \Delta_S = 0.5$. The additional scattering modes make qualitative changes in the behavior of $|r_{h,D}^{A}|^2$ and $|r_{e,B}^{A}|^2$ at small angles of incidence $\alpha_A$ and small changes for large values of $\alpha_A$. The probability of retro type AR from the D point increases by increasing $\alpha_A$ for small angles of incidence and then decreases with $\alpha_A$ and reaches zero for $\alpha_A>\alpha_h^D$. The probability of specular AR from the F point is very small in comparison with that of the retro type AR from the D point and has an increasing behavior with {$\alpha_A$} for $\alpha_A<\alpha_h^F$ and is forbidden for $\alpha_A>\alpha_h^F$. The normal reflection probability from the B point increases with $\alpha_A$ and perfect normal reflection occurs for $\alpha_A>\alpha_h^D$, while the probability of normal reflection from the C point decreases with $\alpha_A$ and goes to zero for $\alpha_A>\alpha_e^C$. Therefore, the dominant AR process is retro type AR, and normal reflection with wave vector $\bm{k}_B$ is the only possible reflection process (with $|r_{e,B}^A|^2=1$) for incident angles $\alpha_A>\alpha_h^D$.

The critical angles of incidence for different reflection processes (above which certain types of reflections are forbidden) can be defined via the conservation of the y-component wave vector $k_y$ under the scattering processes as, $\alpha_e^C=\arcsin(|\bm{k}_C|/|\bm{k}_A|)$, $\alpha_h^D=\arcsin(|\bm{k}_D|/|\bm{k}_A|)$, and $\alpha_h^F=\arcsin(|\bm{k}_F|/|\bm{k}_A|)$, where $\alpha_h^D>\alpha_e^C>\alpha_h^F$. In the inset of Fig. \ref{Fig:3}(a), we demonstrate that increasing the coupling parameter between the two surfaces of the thin film leads to the reduction of the probability of retro AR, $|r_{h,D}^{A}|^2$, and critical angle of incidence, $\alpha_h^D$, the enhancement of the probabilities of normal reflections, $|r_{e,B}^{A}|^2$ and $|r_{e,C}^{A}|^2$, and the critical angles $\alpha_e^C$ and $\alpha_h^F$.
\begin{figure}[t]
\begin{center}
\includegraphics[width=3.4in]{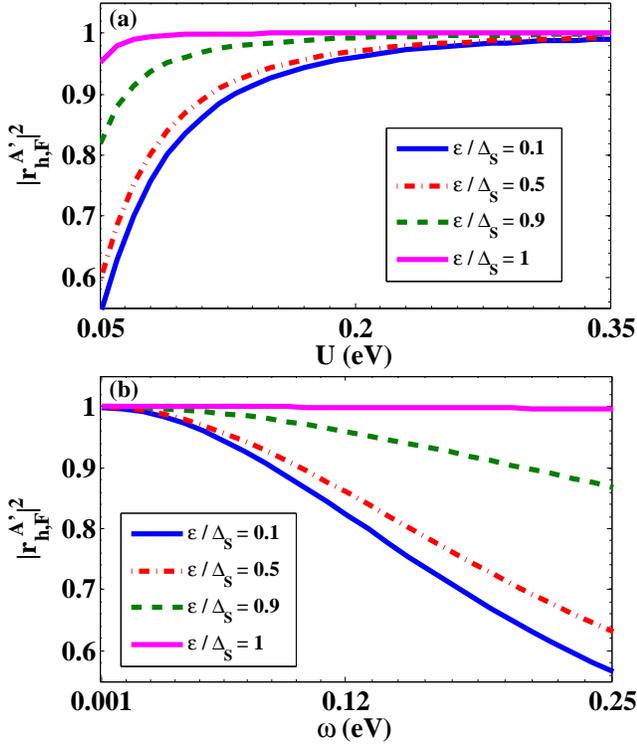}
\end{center}
\caption{\label{Fig:4} Dependence of the specular AR probability for a near-normally incident electron (with $\alpha_{A'}=\pi/90$) from A' point on (a) the top gate potential $U$, when the coupling parameter of the two surface states is $\omega=0.05$ eV, and (b) the coupling parameter $\omega$, when $U=0.2$ eV.}
\end{figure}
\begin{figure}[t]
\begin{center}
\includegraphics[width=3.4in]{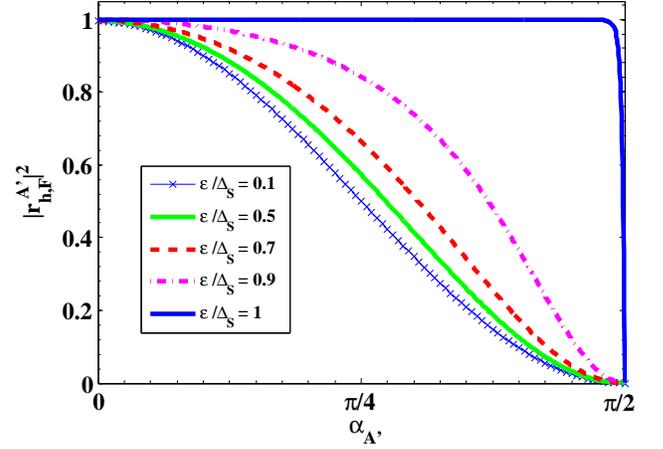}
\end{center}
\caption{\label{Fig:5} Probability of specular AR versus the angle of incidence for an incident electron from the A' point with different values of the excitation energy, when $U=0.4$ eV and $\omega=0.05$ eV. Of particular interest is the perfect specular AR for all angles of incidence to the N/S interface with $\varepsilon=\Delta_S$, when $U$ is large.}
\end{figure}

Figure \ref{Fig:3}(b) shows the behavior of the probability of normal and Andreev reflection processes for an incident electron from the A' point in terms of the angle of incidence $\alpha_{A'}$. Most importantly, the dominant AR process is specular AR, and the contribution of retro type AR to the charge transport is very small. The probabilities of specular AR from the F point, $|r_{h,F}^{A'}|^2$, and of normal reflection from the B point, $|r_{e,B}^{A'}|^2$, decrease with the angle of incidence $\alpha_{A'}$ and go to zero at $\alpha_{A'}=\pi/2$, while the probability of normal reflection from the C point, $|r_{e,C}^{A'}|^2$, increases with $\alpha_{A'}$, and perfect reflection happens at $\alpha_{A'}=\pi/2$. Therefore, we emphasize that the dominant AR process for an incident electron from the A' point to the TI thin-film-based N/S interface is specular AR, while for an incident electron from the A point it is retro type AR. As mentioned before, this specular type AR is intraband in contrast to that of the corresponding structure with uncoupled surfaces ($\omega=0$).

In Fig. \ref{Fig:4}(a), we plot the probability of specular AR for a near-normally incident electron from the A' point (with $\alpha_{A'}=\pi/90$) as a function of the top gate potential, $U$, for four different values of $\varepsilon$, when $\omega=0.05$ eV. The probability of specular AR increases by increasing the strength of the applied gate electric field, and the electron-hole conversion happens with unit probability for $U>U_0$. The value of $U_0$ decreases with increasing excitation energy, such that we have a perfect AR of specular type for a wide range of $U$, when $\varepsilon=\Delta_S$.

Moreover, we present the behavior of the probability of such a specular AR in terms of the value of $\omega$ for different excitation energies, when $U=0.2$ eV [see Fig. \ref{Fig:4}(b)]. For $\varepsilon<\Delta_S$, the probability of specular AR strongly depends on the value of $\omega$ and decreases from unit probability at small values of $\omega$ by increasing the coupling parameter of the two surfaces, which depends on the thickness of the thin film~\cite{zhang10}. In the case of $\varepsilon=\Delta_S$, the probability of the electron-hole conversion in a specular AR does not depend on $\omega$, and perfect specular AR occurs for a wide experimentally available range of $\omega$. Therefore, by tuning the gate-induced potential difference between the top and bottom surfaces of the TI thin film with small values of $\omega$, we will have perfect specular AR for a near-normally incident electron from the A' point with different values of $\varepsilon$, in contrast to that of the corresponding 2DEG structure~\cite{Lv12}, where perfect specular electron-hole conversion occurs in the presence of the strong Rashba spin-orbit interaction ($\lambda=0.4$ eV) for a near-normally incident electron with $\varepsilon=\Delta_S$.
\par
\begin{figure}[]
\begin{center}
\includegraphics[width=3.4in]{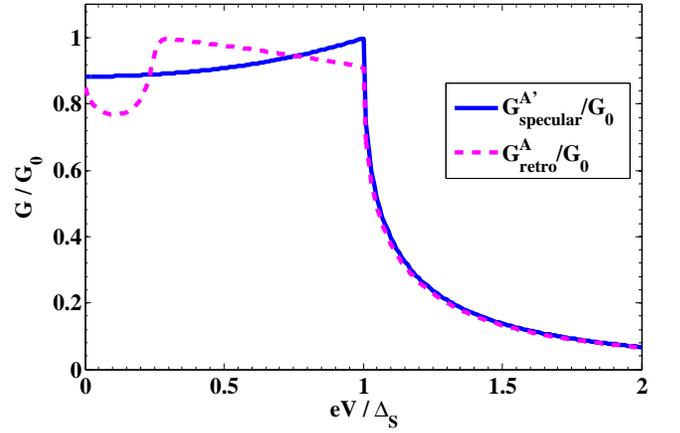}
\end{center}
\caption{\label{Fig:6}Retro and specular Andreev conductances, respectively, for an incident electron from A and A' points to the N/S junction versus the bias voltage $eV/\Delta_S$ (in units of the superconducting gap $\Delta_S$), when $U =0.2$ eV and $\omega=0.05$ eV.}
\end{figure}
\begin{figure}[]
\begin{center}
\includegraphics[width=3.4in]{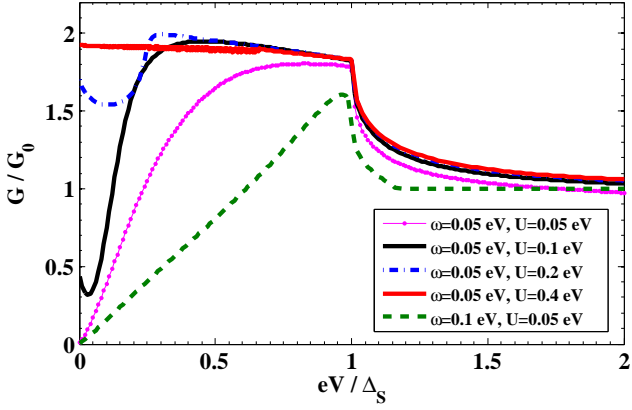}
\end{center}
\caption{\label{Fig:7}Andreev conductance of the N/S junction as a function of the bias voltage $eV/\Delta_S$ for different values of $U$ and $\omega$.}
\end{figure}
Furthermore, the dependence of the specular AR probability of an incident electron from the A' point on the angle of incidence is presented in Fig. \ref{Fig:5} for different values of $\varepsilon$, when $U=0.4$ eV and $\omega=0.05$ eV. Importantly, perfect conversion of an electron to a specular Andreev reflected hole is possible not only in the case of near normal incidence, but also for all angles of incidence to the N/S interface of the proposed structure with $\varepsilon=\Delta_S$, when $U$ is large. This property is another advantage of the proposed TI thin-film-based structure over graphene- and 2DEG-based structures~\cite{beenakker06,beenakker08,Lv12}, where the electron-hole conversion with unit probability only occurs for normal incidence of the electron to the N/S interface of graphene, when $\varepsilon\leq\Delta_S$ and for a 2DEG with strong Rashba spin-orbit coupling ($\lambda=0.4$ eV), when $\varepsilon=\Delta_S$.

As the retro and specular Andreev reflections are respectively the dominant AR processes for incident electrons with the wave vectors $\bm{k}_A$ and $\bm{k}_{A'}$, we plot the bias voltage dependence of the normalized retro Andreev conductance $G_{retro}^A/G_0^A$ and the normalized specular Andreev conductance $G_{specular}^{A'}/G_0^{A'}$ in Fig. \ref{Fig:6}, where $G_{retro(specular)}^{A(A')}={e^2 W}\int_{0}^{|\bm{k}_{A(A')}|}dk_y\ |r_{h,D(F)}^{A(A')}|^2/{2\pi^2\hbar} $, $G_0^{A(A')}=e^2\ W |\bm{k}_{A(A')}|/2\pi^2\hbar$, $U = 0.2$ eV, and $\omega = 0.05$ eV. For $eV/\Delta_S\leq 1$, we find that the specular Andreev conductance shows an increasing behavior with the subgap bias voltage $eV/\Delta_S$ (in units of the superconducting gap $\Delta_S$) and reaches unity at $eV/\Delta_S=1$, while the retro Andreev conductance displays a valley and then a peak at small values of the bias voltage and decreases with $eV/\Delta_S$ for large values of $eV/\Delta_S$. In the case of $eV/\Delta_S>1$, both specular and retro Andreev conductances decrease with increasing $eV/\Delta_S$ value. Since the experimentally measurable Andreev conductance contains the contributions from all transport channels, we present the behavior of the normalized total Andreev conductance of the N/S structure $G/G_0$ in terms of the bias voltage $eV/\Delta_S$ in Fig. \ref{Fig:7} for different values of $U$ and $\omega$, where $G_0=e^2\ W (|\bm{k}_A|+|\bm{k}_{A'}|)/2\pi^2\hbar$ is the normal state conductance of a sheet of TI thin film of width $W$. It is seen that the Andreev conductance has a singularity at $eV/\Delta_S=1$ and a decreasing behavior with the bias voltage for $eV/\Delta_S>1$, as usual for an N/S junction. In the case of $U=\omega=0.05$ eV, the Andreev conductance increases monotonically with the subgap bias voltage and reaches a limiting maximum value for large values of $eV/\Delta_S$. Increasing $U$ leads to the appearance of a valley at small values of $eV/\Delta_S$, which moves towards larger bias voltages by increasing $U$. The enhancing subgap Andreev conductance tends toward the nearly constant Andreev conductance of the corresponding N/S structure with large U. However, if we enhance the value of $\omega$, the Andreev conductance will attain a maximum around $eV/\Delta_S=1$.
\begin{figure}[t]
\begin{center}
\includegraphics[width=3.4in]{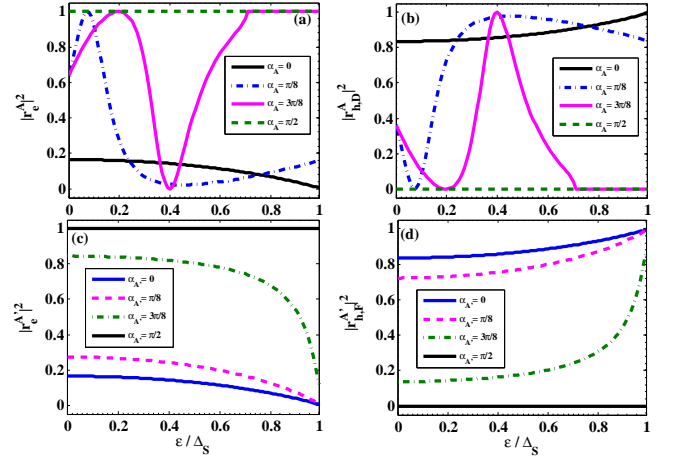}
\end{center}
\caption{\label{Fig:8} The energy dependence of the normal (left panel) and Andreev (right panel) reflection probabilities for different angles of incidence of the electron to the N/S interface from A [(a), and (b)] and A' [(c), and (d)] points, when $U=0.1$ eV and $\omega=0.05$ eV.}
\end{figure}

In order to explain the behavior of the subgap Andreev conductance, we plot the energy dependence of the normal (left panel of Fig. \ref{Fig:8}) and Andreev (right panel of Fig. \ref{Fig:8}) reflection probabilities in Fig. \ref{Fig:8} for the incident electron from A [Figs. \ref{Fig:8}(a) and \ref{Fig:8}(b)] and A' [Figs. \ref{Fig:8}(c) and \ref{Fig:8}(d)] points with four different angles of incidence, when $U=0.1$ eV and $\omega=0.05$ eV. The contribution of an incident electron from the A point to the Andreev conductance is of retro type AR (from the D point) with the appearance of a minimum and then a maximum by increasing $\varepsilon/\Delta_S$. Specular type AR (from the F point) of an incident electron from the A' point leads to increasing behavior of the Andreev conductance with $\varepsilon/\Delta_S$.

\begin{figure}[t]
\begin{center}
\includegraphics[width=3.4in]{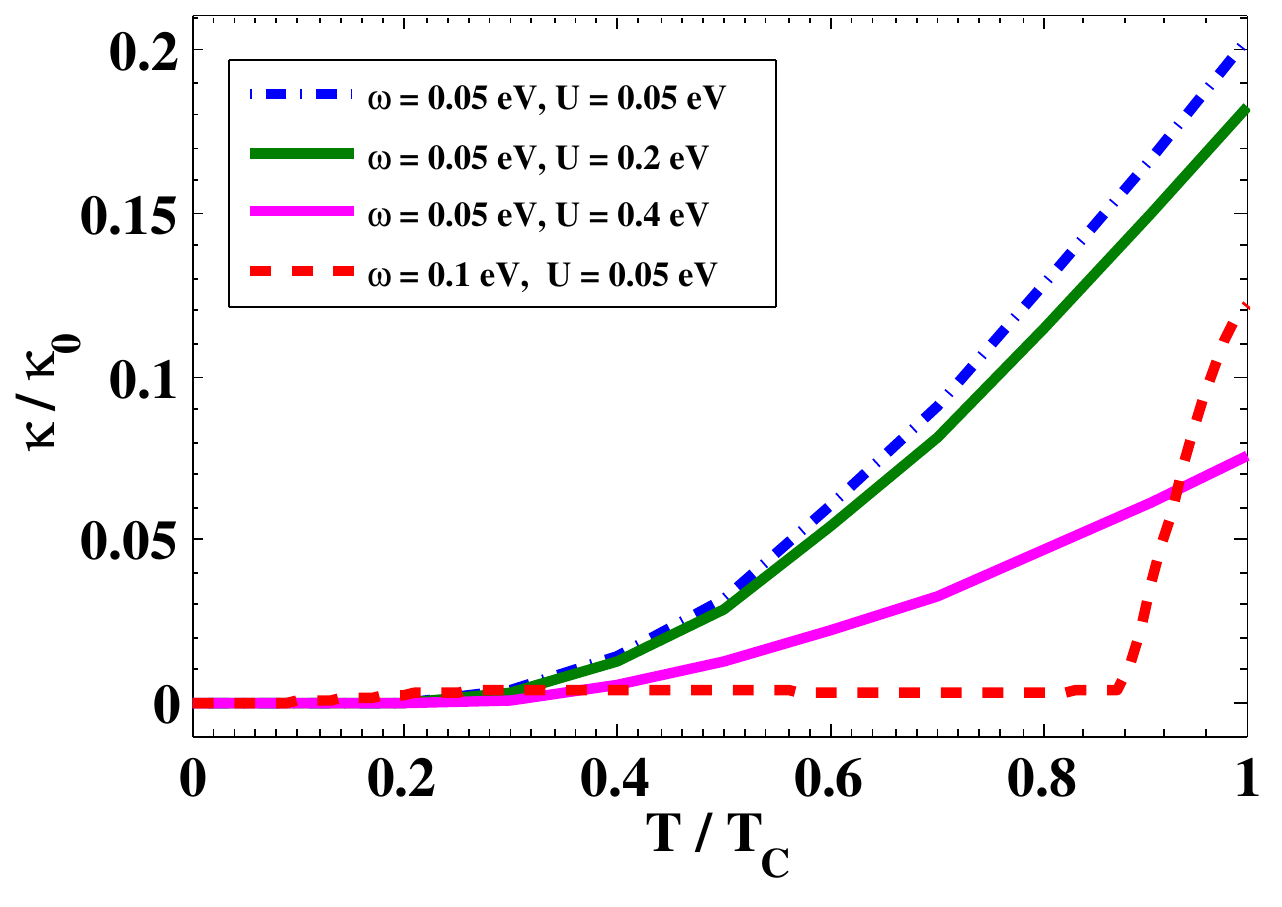}
\end{center}
\caption{\label{Fig:9}Thermal conductance of the N/S junction as a function of the temperature $T/T_C$ (in units of the critical temperature of the superconducting order parameter $\Delta_S$) for different values of $U$ and $\omega$.}
\end{figure}

Finally, in order to complete the investigation of the transport properties of the proposed N/S structure, we present the behavior of the normalized thermal conductance $\kappa/\kappa_0$ given by Eq. (\ref{kappa}) [where $\kappa_0={k_B W}\sum_{i=A,A'}\int_0^{\infty}d\varepsilon\ {|\bm{k}_i(\varepsilon)|} /{8\pi^2 \hbar}$] with respect to the temperature $T/T_C$ (in units of the critical temperature of the superconducting order parameter $\Delta_S$) for different values of $U$ and $\omega$ in Fig. \ref{Fig:9}. We find an exponential dependence of the thermal conductance on the temperature, where it vanishes at temperatures well bellow $T_C$. This behavior of the thermal conductance is similar to that of a conventional normal metal/superconductor junction~\cite{Andreev64} and reflects the s-wave symmetry of the superconducting TI thin film. Also, the thermal conductance decreases by enhancing $U$, while it can be decreased or increased by enhancing $\omega$, depending on the value of the temperature $T/T_C$.
\section{\label{sec:level3}Conclusion}
We have investigated the proximity effect in a hybrid structure of a superconducting (S) and normal (N) topological insulator (TI) thin film. We have realized the possibility of specular Andreev reflection (AR) in the presence of a perpendicular gate electric field in such a structure. We have demonstrated that the probability of specular AR increases by enhancing the gate-induced potential difference between the two surfaces of the TI thin film, $2U$, and specular electron-hole conversion with unit efficiency happens for large values of $U$. We have further analyzed the effect of the coupling parameter of the top and bottom surfaces, $\omega$, and demonstrated that the probability of specular AR decreases with $\omega$ for near normal incidence of electrons to the N/S interface with excitation energies smaller than the superconducting gap ($\varepsilon<\Delta_S$), while unit specular electron-hole conversion is possible for a wide experimentally available range of $\omega$, when $\varepsilon=\Delta_S$. Furthermore, we have shown that perfect specular AR happens for all angles of incidence to the N/S interface with $\varepsilon=\Delta_S$ and large values of $U$, in contrast to the corresponding structures with graphene and two-dimensional electron gas with strong Rashba spin-orbit interaction. We claim that our proposed TI thin-film-based N/S hybrid structure may be a suitable experimental structure for the realization of specular AR; the practical significance of a graphene-based structure rests on the fabrication of high quality samples, and the corresponding structure with a two-dimensional electron gas needs a strong Rashba spin-orbit interaction.

Moreover, we have demonstrated that the Andreev differential conductance of the N/S junction has an increasing behavior with the subgap bias voltage $eV/\Delta_S$ for small values of $U$ and $\omega$, when $U=\omega$. The enhancement of the potential difference $2U$ between the two surfaces leads to the appearance of a minimum value at small $eV/\Delta_S$, while by increasing $\omega$ it attains a maximum value at large subgap $eV/\Delta_S$.

To complete the investigation of the transport properties of the device, we have further presented the thermal conductance of the proposed N/S structure. We have shown that the thermal conductance exhibits exponential dependence on the temperature, which reflects the s-wave symmetry of the superconducting TI thin film. Depending on the value of the temperature, the thermal conductance can be increased or decreased by increasing $U$, while it decreases by increasing $\omega$. Our specular AR can be explored by experiment in the proposed topological insulator thin-film-based N/S structure.
\section{Acknowledgments}
This work is partially supported by Iran Science Elites Federation.

\end{document}